# Planar Superconductor-Insulator-Superconductor Mixer Array Receivers for Wide Field of View Astronomical Observation

W. Shan*[a,b], S. Ezaki [a], J. Liu[c], S. Asayama [a], T. Noguchi [a,b], and S. Iguchi [a,b]

[a] National Astronomical Observatory of Japan, Osawa 2-21-1, Mitaka, Tokyo, Japan, 181-8588; [b] SOKENDAI (The Graduate University for Advanced Studies), Shonan Village, Hayama, Kanagawa, Japan 240-0193; [c]Purple Mountain Observatory, Chinese Academy of Sciences, 8 Yuanhua Road, Nanjing, China 210034

## ABSTRACT

We present a conceptual framework of planar SIS mixer array receivers and the studies on the required techniques. This concept features membrane-based on-chip waveguide probes and a quasi-two-dimensional local-oscillator distribution waveguide network. This concept allows sophisticated functions, such as dual-polarization, balanced mixing and sideband separation, easily implemented with the SIS mixer array in the same planar circuit. We have developed a single-pixel prototype receiver by implementing the concept in the design. Initial measurement results show good evidences that support the feasibility of the concept.

## 1. INTRODUCTION

Extended sources in millimeter and sub-millimeter astronomical spectral line observation, such as galactic star formation regions, nearby galaxies, Magellanic Clouds, and solar observations, require intensive observation time to image and increase the completion in observation resources.  It is, therefore, a natural trend of efforts to expand the instantaneous field of view (FoV) by using of heterodyne arrays at the focal plane. In cosmological distance, even though the objects are not resolved, wide FoV benefits observations in blind statics of gas in high-z galaxies, which suffers a lot by low efficiency with single beam receivers. Currently, many Atacama Large Millimeter/submillimeter Array (ALMA) fields require multiple pointings for achieving their science goals. Modest-size arrays are considered to be an important future development for the whole array or for a subset of antennas, for instance the four in the Total Power array [1, 2]. Wide FOV will also be important in terms of synergetic studies with future instruments such as Thirty Metre Telescope (TMT).

Focal plane arrays at millimeter and sub-millimeter wavelengths are challenging instruments. While direct detection cameras with thousands of pixels have been developed, large-format multipixel heterodyne receivers are completely absent because of enormous technical difficulties [3]. Due to low noise performance approaching the quantum limit, Superconductor-Insulator-Superconductor (SIS) tunnel junctions are dominantly used for ground-based telescopes. Most of the existing SIS array frontends are assembled by putting together individual building block of single-pixel modules. Higher level integration with one-dimensional arrays has achieved the maximum pixel count of 64 [4]. Large number of pixels are extremely difficult if the mixers are configured to perform side-band separation or dual polarization receiving due to the increase of the complexity by involving more waveguide components, such as waveguide hybrid bridges and waveguide orthomode transducer (OMT) [5-6].  An alternative solution for dual polarization SIS arrays is to make a duplicate array for another polarization as done in [7] with a cost of doubling the system scale.

Focused on the frontend receiver technology, the major reason that causes the difficulties in building a large format array lies in the distribution network of local oscillator (LO) and LO/signal diplexers. The necessity of reference signal is the essential difference from the direct detection cameras with superconducting detectors (MKID or TES). Either in the form of metal waveguides or quasi-optical beam splitters, the LO distribution network and the LO/signal diplexers contain complicated three-dimensional structures, which are difficult to be packed into a compact array. Another difficulty that prevents a large-format array lies in the mixer chip. The conventional waveguide SIS mixers based on, for example, quartz substrate, couple the LO and signal through a waveguide probe. This approach brings about two difficulties in making an array. First, it prohibits the use of a monolithic planar chip that serves for all pixels like in the direct detection cameras. Second, a waveguide LO distribution that allows the LO/signal diplexers with waveguide couplers is not practical with convention machining unless high performance additive manufacturing method can be applied. The metal additive manufacturing is still in the early stage even though rapid progress has been made [8].

*wenlei.shan@nao.ac.jp; phone 81 422 34 3864

The aim of this study is to reform the integration method to allow a compact array and possible extension to large pixel counts. To achieve high compactness, several functional layers that serve all pixels in a 2D array should be the fundamental building blocks. In this way, mechanical interconnection between pixels is inherently fulfilled. In comparison, these interconnections bring about bulk size and uncertainties in a conventional SIS mixer array formed with separated pixel modules. To implement this planar integration concept into LO distribution is difficult if LO/signal diplexers are put into the LO distribution layer. This difficulty can be solved if the LO/signal diplexers is moved to the mixer chip, with only the hierarchical tree-structure LO distribution network remaining in the LO distribution layer. The mixer chip, no longer a carrier for a single pixel but for all of the pixels, contains antennas that couple LO and signal independently. The transmission of LO and signal from the waveguides to the planar circuit is achieved through membrane-based probes fabricated on locally thinned substrates. Silicon on insulator (SOI) wafers and deep reactive ion etching are employed to involve the membrane technique in the conventional SIS junction fabrication process. The use of membrane-based LO probes disentangles the mixer IC from the LO distribution network, and therefore, there is no restriction in the geometry of the IC. This allows us to incorporate many sophisticated capabilities, such as dual-polarization, sideband separation, and balanced mixing, into an SIS mixer array, which is not practical by assembling single-pixel modules in case the pixel count rises.

In this paper, the concept of planar integration of SIS mixer array is framed in section 2. It is followed by a proof-of-concept study, in which the concept is implemented in a dual polarization single-pixel SIS mixer design at 2 mm wavelength with a balanced mixing configuration. In section 3, the integrated planer circuit design and the mixer block design are introduced in details. In section 4, the performance of the prototype receiver is evaluated. The SIS mixer performance and performance of the planar OMT were measured.

In parallel with the SIS mixer development, we are conducting studies to develop cryogenic low-power consumption low-noise amplifiers (LNAs) at intermediate frequency (IF). The power consumption of LNA will become a limiting factor of SIS array frontend if the pixel count rises. We have been developing superconducting microstrip line parametric amplifiers, which reveal a several-dB parametric amplification at 6 GHz band and at an ambient temperature of 4 K [9]. A GaAs MMIC has been designed to operate at 2-4GHz band. Initial measurement results demonstrate a 20 dB gain at 4 K with DC power of 1 mW [10].

## 2. THE PLANAR INTEGRATION CONCEPT

### 2.1 Layered Structure and Metal Waveguide LO Distribution Network

In the proposed planar integration concept, the assembly is composed of several functional layers. The essential idea of the proposed integration concept lies in that the monolithic functional layers serve for all pixels. Unlike in the traditional approach where individual single-pixel modules are connected to form an array, in this approach no mechanical interconnection is needed between pixels. This greatly simplifies the architecture and enables very compact integration.

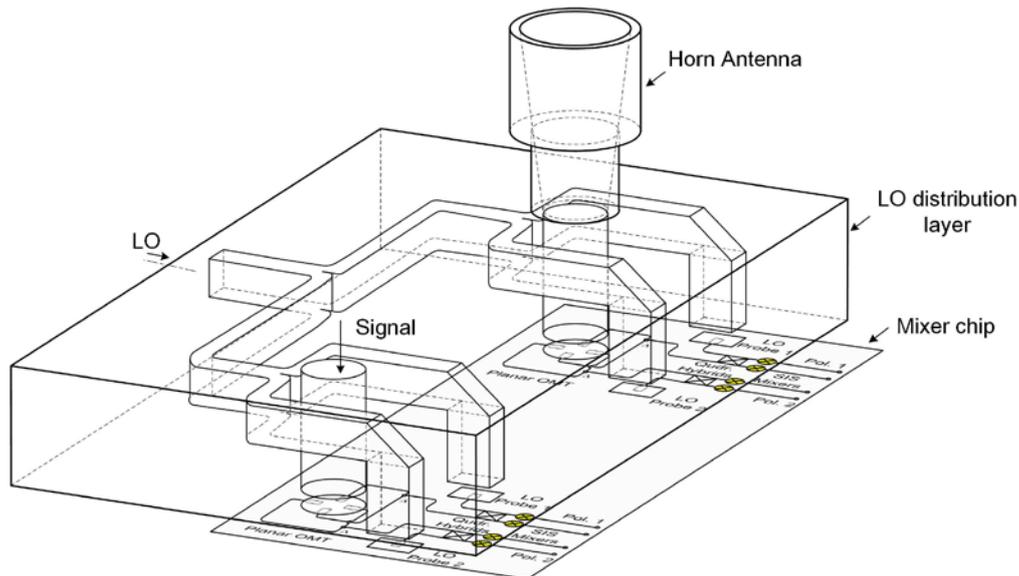

Figure 1. Schematic drawing of a two-pixel array to explain the planar integration concept with the LO distribution layer and the mixer chip emphasized. The equivalent circuit shown in this plot includes two identical dual-polarization balanced SIS mixers. A single-pixel model has been experimentally tested in this study.

A schematic drawing illustrating the above mentioned concept is shown in Figure 1. Without loss of generality, a two-pixel array is used to explain all the key features. The top layer is a horn antenna array. For a clear graphic explanation, this layer is not drawn to be a metal block with embedded horn antennas though more realistic in practice. Instead, for one pixel an individual horn antenna is attached and for another the antenna is absent to show the signal waveguide. The layer that follows is the LO distribution layer, in which a simple one-to-four power divider is configured as a tree-structure network spreading in a horizontal plane. The signal waveguides go vertically through this layer are physically isolated from the LO waveguides. The LO waveguides bend 90-degree downward at the ends to guide the LO to the mixer chip layer below. The LO distribution layer is intentionally designed to be simple, not involving any complicated functions, such as waveguide orthomode transducers (OMTs) and LO/signal diplexers, so that they can be machined with high accuracy in a conventional way. The third layer is the mixer chip. In Figure 1, it contains two isolated pixels on the same chip. For each of the pixels, it is configured as balanced mixing as explained later. There are four planar antennas for LO and signal coupling from the waveguide orifices of the LO distribution layer. Backshort cavities are usually necessary to enhance the performance of the LO and signal planar antenna. They are arranged in a forth layer, which is not drawn here.

## 2.2 On-chip Antennas and Planar Circuit

The mixer chip turns out to be complicated because it performs many functions that are conventionally carried out with waveguide components. In this concept, those functions are moved into the chip with planar circuit components. Figure 2(a) shows the photo of the mixer chip for the concept-proof study, which can be regarded as one of the pixels shown in Figure 1.

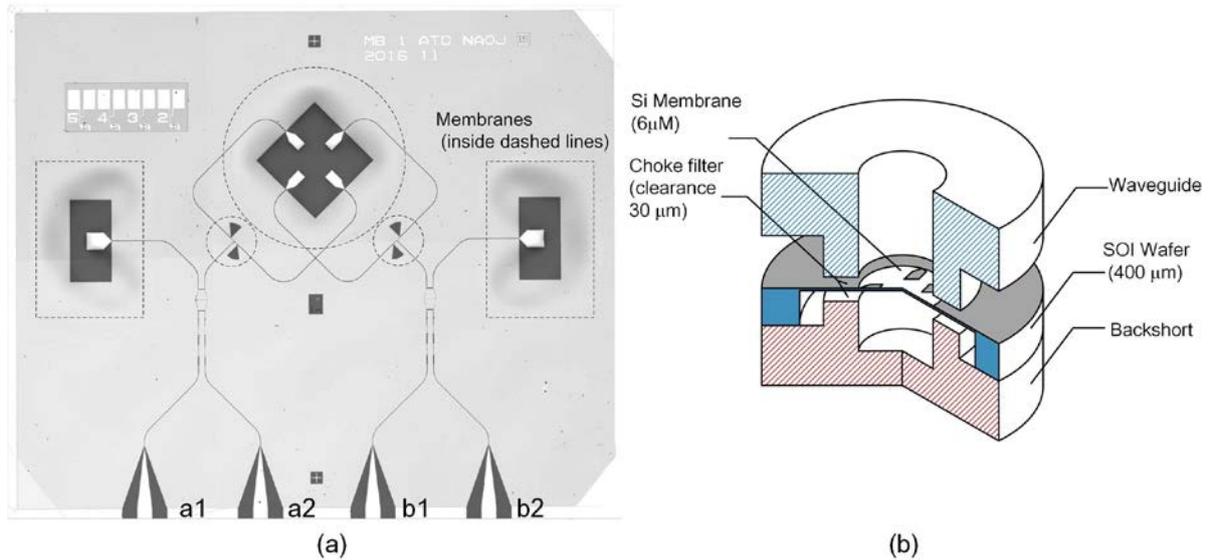

Figure 2. (a) The photograph of the planar integrated circuit for the concept-proof study. The dashed lines indicate the borders of the membrane area. (b) The schematic plot of the membrane base planar antenna for the signal coupling.

Distinct from a conventional SIS mixer on a small piece of substrate that is usually narrower than the broad side of the waveguide, this design is made on a large-area substrate with planar antennas on it. To enable a broadband performance, the substrate under a planar antenna is locally thinned from the backside of the substrate until a membrane remains. The membrane-based waveguide probes have been applied in direct detection cameras in our previous work [11]. Figure 2(b) shows the schematic of the membrane-base probe with some geometric dimensions used in the concept-proof model. The 6 μm Si membrane intervenes between the two waveguide pieces with clearances of 30 micrometers on both sides. The clearance has to be kept small to avoid signal leakage to the outside of the waveguide. To improve the RF sealing, a quarter-wavelength radial waveguide is used as a choke.

Monolithic planar circuits that contain waveguide antenna and LO/signal diplexers have already been used in SIS mixer development [12-14]. Compared with the conventional SIS mixer designs on small chips, these planar circuits can be regarded as system on-chip (SoC). These efforts in developing SIS mixer SoC greatly benefit this study by providing valuable experiences in the design of the planar waveguide components, which are applied in this new concept. This new concept, in particular the membrane-base probes and the LO distribution method, enables a single pixel SoC to be expanded to an array. The uniformity and reliability of a SoC is exactly what an array receiver must have.

## 3. PROOF-OF-CONCEPT MODEL DESIGN

### 3.1 System overview

We have been carrying out a concept-proof study to demonstrate the feasibility of the concept of planar integration of SIS mixer array with a prototype. Although it is a single-pixel SIS mixer, it includes all the key features applicable to a large array. The mixer is configured to be a dual polarization and balanced mixing at 2 mm wavelength. The details of the design are introduced in the following parts of this section. The image of the mixer circuit is shown in Figure 2 and the corresponding circuit diagram is shown in Figure 3. There are two identical sets of balanced mixers with each for one polarization. The planar OMT separates the signal in different polarization by two pairs of probes orientated perpendicularly. The two probes in the same polarization couple signal equally in amplitude but with 180-degree phase difference. They are then combined with an anti-phase power combiner. Unlike the planar OMT, which couples signal of both polarizations, the LO coupling for the two polarizations is separately done with two individual membrane-based probes. An alternative approach is coupling LO with one membrane-based probe and dividing the LO on chip for the two polarization. We chose the former because the metal waveguide power divider can be fabricated with more predictable accuracy. The LO and signal are cross-coupled through a 3-dB quadrature hybrid bridge coupler before being delivered to SIS mixers. The DC-block capacitors that allow RF signal to go through but block the IF are used to avoid adding shunt capacitance from the forepart of the circuit to the SIS mixers, which may limit the instantaneous bandwidth. In the cryogenic IF circuit board, the mixer outputs are cross-coupled in a 180-degree hybrid bridge to reconstruct the signal and separate the LO noise.

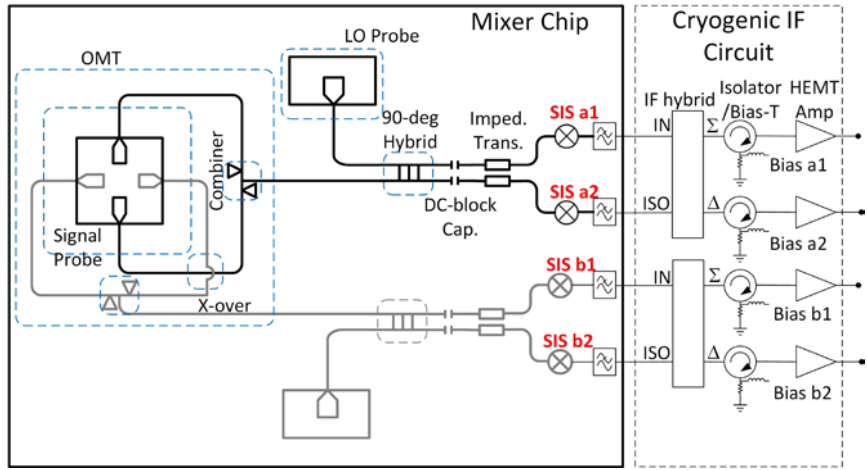

Figure 3. The circuit diagram of the mixer chip for the proof-of-concept study. The cryogenic IF circuit diagram is also given to present a complete balanced mixing scheme.

### 3.2 Thin-film superconducting transmission lines

Two kinds of thin-film superconducting transmission lines are employed in the integrated circuit. For low impedance lines, microstrip (MS) is used with a typical characteristic impedance of about 20 Ω. For high impedance lines, coplanar waveguide (CPW) is used with a typical characteristic impedance of 70Ω. In many places in the circuit, lines with intermediate impedance are desired. In those places electorally-short sections of CPW and MS are connected alternatively to form an equivalent transmission line with desired characteristic impedance as introduced in [12]. This flexibility in the selection of the characteristic impedance of a transmission facilitates the designs of the planar waveguide components such as couplers and impedance transformers, and particularly mitigates the difficulties in the designs of those components with broadband performance.

To keep the same potential of the ground planes in a CPW configuration, underpasses connecting the two ground planes are used as shown in Figure 4(a). The distance between two underpasses is set to be about quarter wavelength at center frequency, keeping the lowest trapped oscillation mode far beyond the operating frequency range. In this study, the underpasses and the center strip are separated by a thin silicon dioxide layer of a thickness of 300 nm. The width of the underpass is 2 μm. Since the width of the underpasses is much shorter than the wavelength at 2 mm wavelength, it does not bring about noticeable changes to the characteristic impedance of the CPW. At much operating higher frequency (sub-mm), air-bridges formed by a photolithography method will be a better solution.

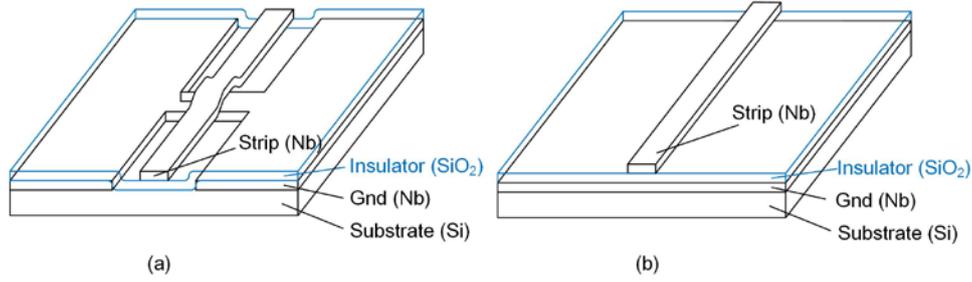

Figure 4. (a) The configuration of the CPW with an underpass connecting the ground planes. The silicon dioxide layer is 300 nm, Nb thickness for ground plane and strip are 300 nm and 400 nm respectively. Typically, the strip width is 3 μm and the gap between the central strip and the ground plane is 4 μm. The above structure leads to a characteristic impedance to be about 63 Ω and wavelength to be about 755 μm with taking the kinetic inductance of the Nb strip into consideration; (b) the structure of the MS. The insulator and superconductor thickness are the same as in CPW. For a strip width 3 μm, the characteristic impedance of the MS turns to be about 19 Ω and the wavelength about 900 μm.

At 4 K ambient temperature, the RF loss of a Niobium (Nb) superconducting transmission line is not negligible because of the existence of thermal agitated quasi-particles. The loss that takes place in the signal transmission pass before the SIS mixer will cause increase in the equivalent noise of the SIS receiver. We calculated the transmission loss based on Matthias-Bardeen theory. The overall lengths of CPW and MS in front of the SIS mixer in the chip are about 4.4 wavelengths and 1 wavelength respectively at 145 GHz. The losses are 0.03 dB and 0.06 dB per wavelength respectively. This results in an overall loss about 0.2 dB, equivalent to an increase of equivalent noise about 5%.

### 3.3 On-chip Probes for Signal Coupling and LO

A dual-polarization waveguide probe is designed for signal coupling. The planar OMT design is developed from our previous work [15], with the details shown in Figure 5. The on-chip waveguide probes are carried by a 6 μm thick silicon membrane, which is sandwiched by two mental waveguide pieces with a narrow gap above and below the membrane. The cross section of the signal waveguide shapes into square with the side length equaling the broad side of the WR-6 waveguide in order to match a corrugated horn, which we already have in the lab. An OMT that fits a circular waveguide has also been designed. The performance does not show much difference.

The membrane breaks the RF current in the waveguide sidewall and causes the signal leaking outside the waveguide. To minimize this effect, the distance between the two metal pieces on both sides of the membrane is designed to be close to each other to form a capacitor which can bypass the RF current effectively. This structure can be also regarded as a radial waveguide with a low characteristic impedance. Therefore, a quarter-wavelength choke filter can be designed to enhance the seal of the waveguide preventing the RF signal from leaking out. We chose a clearance of 30 μm, the maximum value that allows a frequency coverage of the whole WR-6. A smaller value is not chosen because we want to reserve the possibility to down-scale this design for higher frequency operation. Taking into consideration the limit of the machining accuracy, a clearance of about 10 μm is still practical. If less than this value, there will be a risk of the damage in the membrane caused by physical contact to the mixer block pieces. Given a clearance of 10 μm, this design can be scaled down by a factor of 3. It means that it can be used up to about 500 GHz, supposing that there is no other limiting factors. For even higher frequencies, more precise machining, for example, silicon micromachining has to be considered.

Figure 5(b) shows an ANSYS HFSS® simulation result of E filed distribution on the surface of the front side of the membrane at 145 GHz. Two probes of the same polarization are driven by two sources respectively with same amplitude but in 180-degree phase difference. The simulation result shows some field appearing in the choke region but nothing outside. Under this antisymmetric driven condition, an E-field null forms at the place where the other pair of probes locate. A perfect isolation can be achieved between two polarizations if the symmetry is precisely hold. In practice, cross talk happens if the symmetry is lost because of miss alignment of the chip, machining errors, or cross-talk elsewhere, such as at the cross-over. This probe is optimized to have a broadband performance as shown in Figure 4d. An input impedance of about 100 Ω is almost frequency-independent . The impedance is calculated with supposing that there is a perfect anti-phase combiner that combines the signals from the two ports and the impedance is measured at the output port of the combiner. The combined reflection coefficient is $\Gamma = S_{11} - S_{12}$ and the impedance is

$Z_a = Z_0(1-\Gamma)/(1+\Gamma)$, where $Z_0$ is the characteristic impedance of the CPW that connects to the probe, and the S parameters are calculated from the HFSS model.

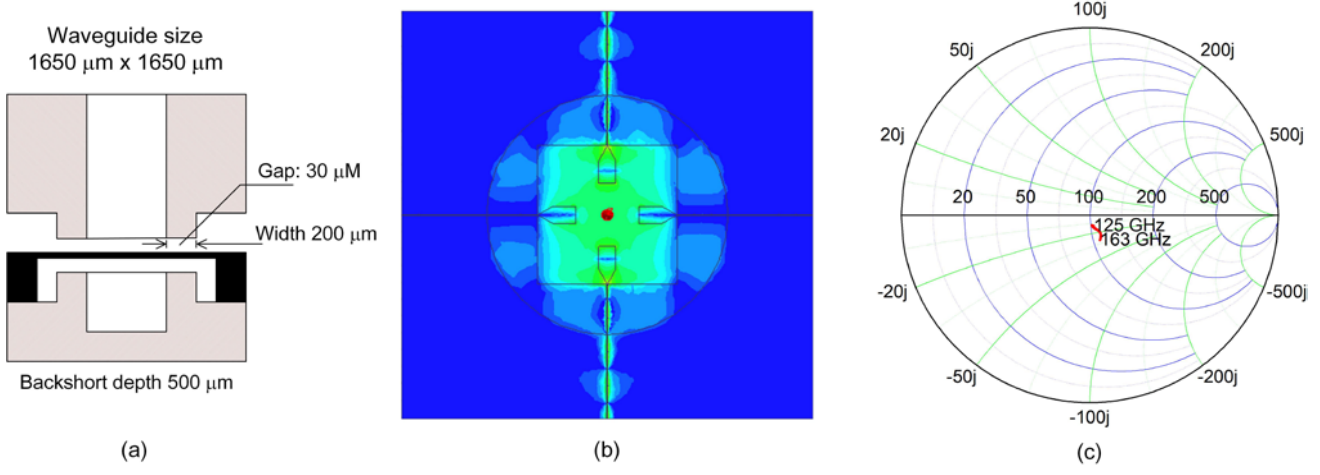

Figure 5. The design of the planar OMT. (a) The schematic plot showing the OMT assembly; (b) the HFSS simulation result of the E filed distribution on the front surface of the OMT at 145 GHz; (c) the impedance of the probe in a frequency range of 125 GHz -163 GHz, otherwise known as ALMA Band 4.

The structure of the LO probe is similar to that of the signal probe with the difference in waveguide cross section and probe configuration. The LO waveguide is a standard WR-6 rectangular waveguide and the probe is single-ended, as shown in Figure 6. Roughly speaking, from the EM simulation point of view the configuration shown in Figure 6(b) is half the model in Figure 5(b). If we put a symmetric E plane along the E-field null in the center of Figure 5(b), the planar OMT becomes the configuration of single-ended probe in Figure 6(b). Consequently, it is not surprising that the impedance is also similar, as shown in Figure 6(c). From figure 6(b) it seems that more field exists in the choke structure. The probe impedance also shows slightly stronger frequency-dependence than the planer OMT does.

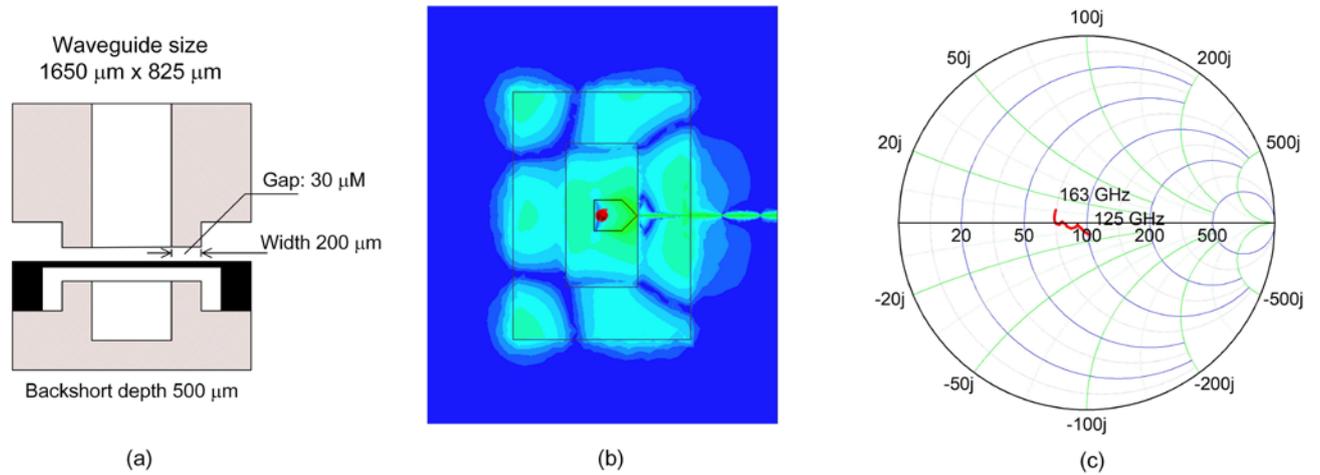

Figure 6. The design of the LO probe. (a) the top-view image of the probe; (b) the schematic plot showing the probe assembly; (c) the HFSS simulation result of the E filed distribution at the plane just above the front surface of the planar antenna at 145 GHz; (d) the impedance of the antenna in frequencies ranging from 125 GHz to 163 GHz.

## 3.4 Anti-phase Combiner and Cross-over

The two anti-phase combiners, each of which combines the signals coupled by a pair of probes in the OMT in one polarization, are slot line-CPW tee-junctions, which are borrowed from [15]. This design has the merit that the anti-phase relation between the two input ports is guaranteed by the geometric symmetry of the circuit, and therefore does not depend on the frequency. The frequency-independent anti-phase relation is also kept between the pair of probes for one polarization in the planar OMT. The layout of the anti-phase combiner is shown in Figure 7 (a), together with the impedance values seen at some specified reference points. For fabrication convenience, the RF grounding of the central strip of the CPW is made with a large-size tunnel junction by utilizing its large capacitance. The large-area junction is designed to have a shape of radial stub with a radius about quarter wavelength of the electromagnetic wave in the barrier to pin down the virtual ground at the joint. It is worth mentioning that the anti-phase combiners are also laid on membranes. It is because HFSS simulation results show that substrate modes in the thick silicon substrate under the slot lines cause significant signal dissipation. However, on a membrane, the loss is negligible.

The unavoidable cross-junction of two signal lines for different polarizations in the planar OMT circuit contributes undesired cross-talk between polarizations. To reduce the coupling from the capacitance in the overlapped area of the two CPWs, the central strips of both lines are tapered as shown in Figure 7(b). Both of the central strips shrink from 4 μm to 1.5 μm at the cross. The insulator (about 300 nm silicon dioxide) is sandwiched by the two lines form a capacitor with a specific capacitance of $0.12 fF/\mu m^2$. The coupling between the two transmission lines through this capacitance is estimated to be less than 30 dB at 2 mm wavelength, with a simulation result shown in Figure 7(b).

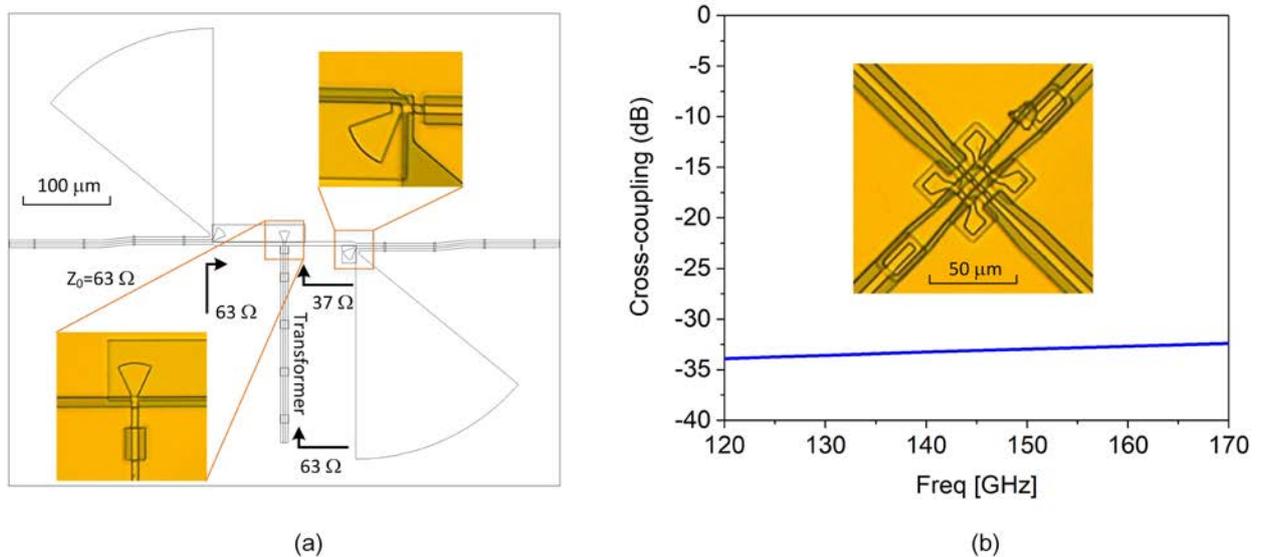

Figure 7. (a) The design of the anti-phase power combiner, and (b) the design and the cross talk calculation result of the CPW cross-over.

## 3.5 Quadrature Hybrid Bridge

The quadrature hybrid bridge is a necessary component in the SIS mixer circuit performing sideband separation or balanced mixing. There are various implements, among which the branch-line type is the most convenient one and therefore most commonly used in SIS planar circuits. In this work a branch-line quadrature hybrid bridge is designed as shown in Figure 8. The transmission line sections with low characteristic impedance (L2 and H2) are in the form of MS. H1 section with the highest characteristic impedance is in the form of CPW. L1 section, which has an intermediate characteristic impedance, is difficult to be implemented with a signal form of planar transmission line, and therefore a synthesized one with a combination of MS and CPW is adopted. From the simulation results, it is found that the characteristic impedance of L2 influence the balance of the hybrid more sensitively than others. About 10% change in the linewidth will cause 1 dB amplitude unbalance in the two outputs. Because of (1) the requirement of the precise characteristic impedance of L2 and (2) the existence of about 0.2 μm uncertainty in the fabrication process, the linewidth of the L2 is designed to be wide enough to ensure a relative uncertainty in linewidth less than 10%. The detailed dimensions for the transmission lines are listed in Table 1. The uncertainty in the thickness of the insulator also causes the deviation of the characteristic impedances from designed values, which is, however, difficult to be controlled in the design.

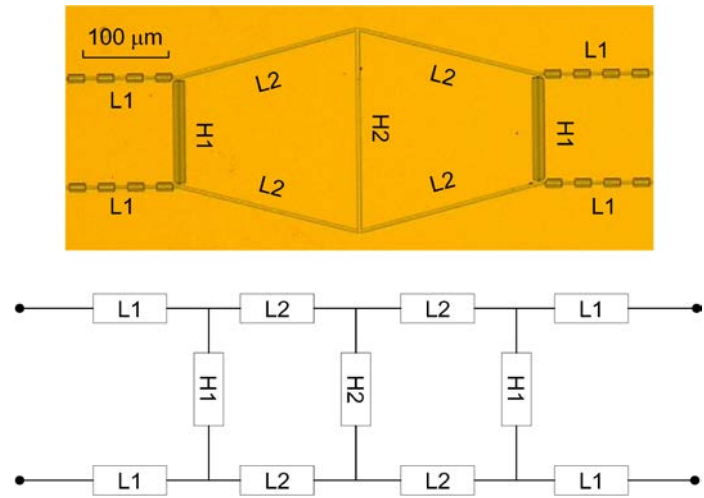

Figure 8. The design of a branch-line quadrature hybrid coupler for balanced mixing.

Table 1. The design parameters for the branch-line coupler shown in Figure 8.

| Transmission line sections | Char. Imp. (Ω) | E-length (deg at 145 GHz) | CPW (MS) Config. (μm) | Physical Length high/low(μm) |
|---|---|---|---|---|
| L1 | 41.4 | 73.8 | 4/3/4 | 20.7/14.3 |
| L2 | 17.1 | 89.0 | 3.6 | 223 |
| H1 | 81.0 | 59.4 | 6.5/2/6.5 | 124.7 |
| H2 | 14.8 | 99.5 | 4.3 | 245 |

### 3.6 SIS Mixers and Tuning Circuits

SIS mixers operating at millimeter frequency ranges easily show high conversion efficiency and even conversion gain can be observed. But this high conversion efficiency is not always favorable because it may cause instability in the DC bias on the pumped IV curve where the dynamic resistance is very large or even negative. Under this circumstance, a small change in bias current can be significantly enlarged in bias voltage due to the large dynamic resistance or even leads to oscillation in bias source. The high conversion efficiency can also cause serious gain compression with the input power from a room temperature blackbody radiation. The gain compression appears because the mixer's IF output drives the bias voltage back and forth in a small range around the DC bias voltage, and thus averages the conversion efficiency in that range. Since this averaging range depends on the signal strength nonlinearity occurs. This nonlinearity will lead to uncertainties in the amplitude calibration of the radio telescope and limits the amplitude accuracy of the scientific data. For ALMA specification, the gain compression by a room temperature blackbody source has to be controlled below 5%. For atmospheric observation, the requirement in linearity is generally higher.

To prevent bias instability and guarantee good linearity, some measures are commonly made in the design phase. The most effective solution is the utilization of an array of SIS junction connected in series [16]. Because the photon steps become wider by a factor of the number of junction in the array, the relative (to the photon step width) averaging range around the DC bias voltage becomes smaller and thus the linearity improves. In this study, we use a 3-juntion series array with junction size of 2 μm in diameter. The three junctions are linked with high impedance lines which tune out the junctions' geometric capacitance, otherwise the nonlinear resistance of the SIS junctions will be bypassed by the linear shunt capacitance and the conversion efficiency will largely reduce. The junction current density is designed to be 8 kA/cm$^2$ to allow a broad-band mixing performance. The junctions are placed electrically close to each other. Therefore, they are driven by the LO almost in phase. Since the phase difference of LO in the array can be neglected, the array can be treated as a single lumped junction as shown in Figure 9(a).

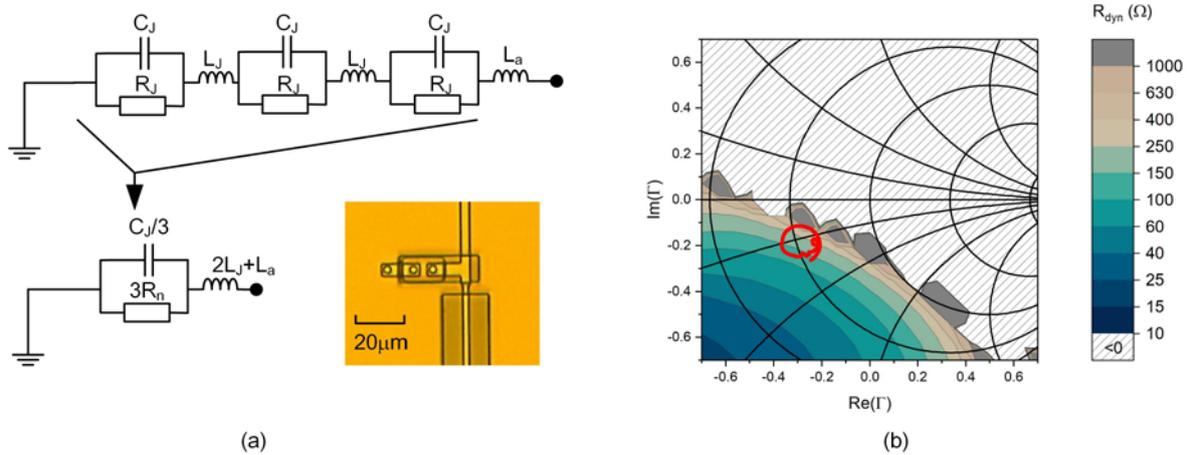

Figure 9. (a) The image of the 3-junction series array and its equivalent circuit; (b) the simulated dynamic resistance as a function of embedding impedance seem by the junction array at 145 GHz (in color keys) and the designed embedding impedance in a frequency range of 125 GHz - 163 GHz.

Another approach for good bias stability and linearity of an SIS is to design the RF impedance seen by the junction to be a little bit capacitive. As shown in Figure 9(b), the dynamic resistance at the DC bias (at the center of the first photon step below the voltage gap) of the junction array is plotted as a function of the embedding impedance at 145 GHz. The embedding impedances in the shaded area induce negative dynamic resistance, which should be avoid in the design. We chose to locate the embedding impedance to the place where the conversion efficiency is moderately high and the dynamic resistance is positive. The simulations were done with a numerical simulation tool, i.e. SISMA (SIS Mixer Analyzer) developed by the authors.

### 3.7 Layered Structure in Mixer Block Design

The mixer block design reflects the major features in the concept that is introduced in this work. As shown in Figure 10, it comprises three slices. The top slice and the middle one compose the complete LO distribution layer. This functional layer is split into two parts to allow the milling of the waveguides buried inside this layer on the upper part. The lower part of the LO distribution layer contains only vertical through-holes (waveguide tubes) that can be machined by wire cutting. The bottom slice serves as the waveguide backshort piece and also as the mixer chip holder. The mixer chip is aligned with three alignment pins that confine the position of the rectangular chips from two perpendicular edges. Because the perimeter of the mixer chip is defined by photolithographic method, the accuracy is better than 1 micrometer. The alignment uncertainty of about ±5μm is mainly caused by the machining errors. Four springs fix the chip from the four corners. To avoid the damage of the membranes due to the difference between the air pressures at the two sides of the membranes during pumping down, three trenches are designed for venting the enclosed air in the chambers under the membranes. A permanent magnet is applied from the bottom of the mixer block for the suppression of the Josephson noise. The IF circuit board is designed to have a broad space to accommodate the IF hybrid bridges.

Figure 11 shows SEM images of the bottom piece of the mixer block at around the signal and LO waveguide backshort cavities. A vent trench can be clearly seen in Figure 11(a). The concave in the center is the backshort cavity, and the plateau around the concave is the lower part of the choke. The membrane should be suspended above the surface of the plateau with a distance of about 30 μm. To verify this, we broke the membrane by a tweezer, and checked it with SEM as shown in Figure 11(b). From this image we confirmed that the gap below the 6 μm silicon membrane is very close to what is expected. From the pattern remaining on the membrane debris, we also confirmed the alignment error of the mixer chip is less than 10 μm.

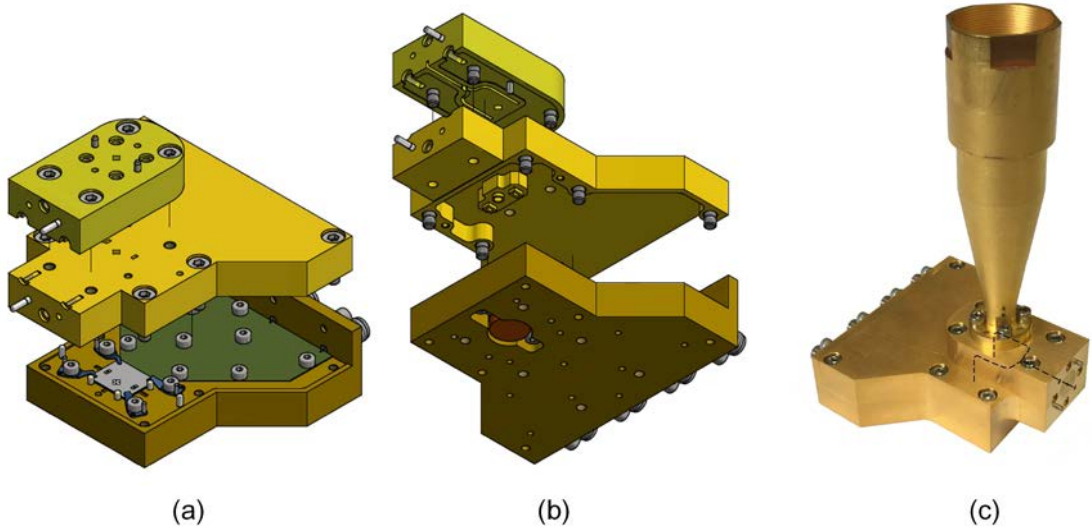

Figure 10. The mixer block design. (a) The exploded view of the CAD model (without horn antenna) from the above perspective; (b) The explode view from below perspective; (c) The photo of machined mixer block with dashed lines indicating the route of LO distribution.

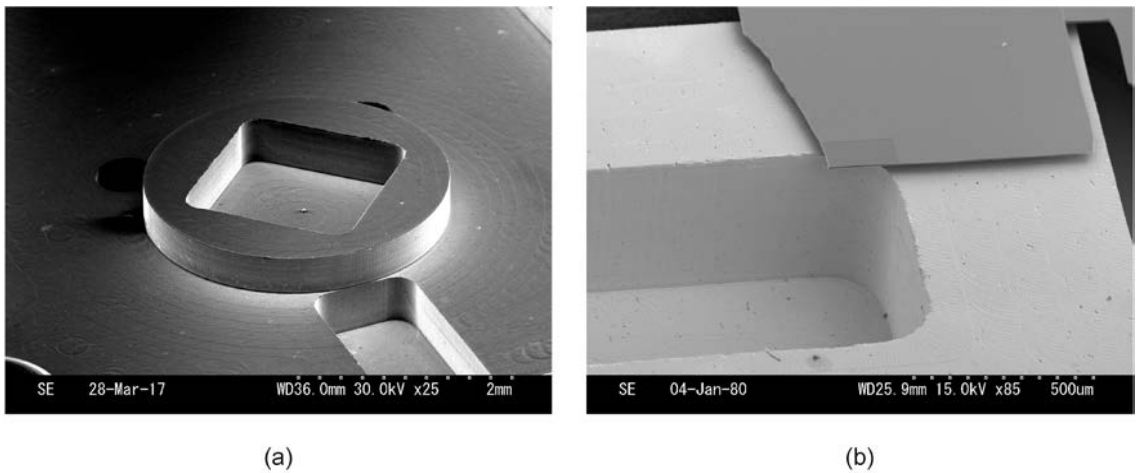

Figure 11. The SEM image of the backshort piece for (a) signal waveguide probe and for (b) the LO waveguide probe. The membrane about the LO waveguide backshort is inventively cracked to show the clearance between the membrane and the metal block.

## 4. EXPERIMENT AND RESULTS

### 4.1 Device Fabrication

The planar-integration approach is fundamentally supported by the SIS mixer fabrication process, which is developed based on our standard SIS mixer fabrication process to enable fabrication membranes in the desired location in a chip. Distinct from a traditional one, this process makes use of SOI wafers and includes a deep reactive ion etching process for removing the handler wafer. The handler wafer, the buried oxidized (BOX) layer and the device layer of the SOI wafer are high-resistance float-zone (FZ) silicon (resistivity at room temperature >5kΩcm), amorphous silicon-dioxide, and FZ silicon with thickness of 400±5 μm, 1μm, and 6μm respectively. The BOX layer serves as an etching stopper when etching the handler where membrane forms. It is removed by applying a supplementary dry etching to minimize the deformation of the membrane caused by its stress. A small but noticeable doming deformation on the final 6 μm-thick silicon membrane persists as shown in Figure 12. This deformation, which is caused by the residue stress in the device layer, is measured to be 0.5% in terms of the ratio of the dome height to the span. This deformation does not bring noteworthy influence on the RF performance according to simulation results.

One thing worth mentioning is that a dedicate $AlO_x$ layer of 50 nm is inserted in the standard process for the DC-block capacitor, which has a specific capacitance of 1.62 fF/μm$^2$. The capacitor size is designed to be 10 μm x10 μm, resulting in a capacitance of 0.16 pF, which will not cause a reflection larger than -20 dB at 145 GHz.

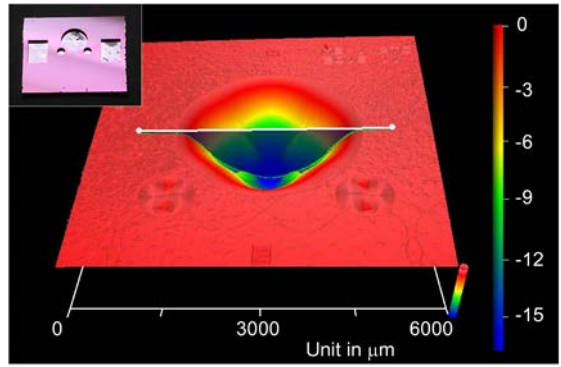

Figure 12. A slight deformation remains in the free-standing membrane, measured with a white light interferometer. The diameter of the circular membrane shown in this image is 3.25 mm, which can be seen from the photo of the backside of the chip shown in the inset.

The mixer chip has an area of 13 mm x 10 mm, which is much larger than a conventional one with a typical lateral dimension of 0.3 mm x 2 mm at the same frequency band. The large size results in a small number of chips per 3-inch wafer that we are using. Unless the fabrication yield is high, the small number of chips per batch does not provide a high chance to get good devices. It is a considerable challenge in our cleanroom facility to improve the fabrication yield to a higher level for large-scale device fabrication, because of the difficulty in clearly identifying and tracing the reasons for the defects founded in the devices. It is obvious that keeping a high level of cleanness is a necessary condition for this device fabrication.

### 4.2 Mixer Performance

Several batches of devices have been fabricated. Due to low yield, only a few devices free from defects can be used in RF measurements. The mixers were evaluated in a 4-K cryostat cooled by a GM cryocooler. The LO is fed through a WR-6 waveguide driven by a synthesized LO source outside the cryostat. A corrugated horn is attached to the mixer block to couple signal through a vacuum window covered by a thin polyimide film. The mixer noise is measured using the standard Y-factor method with two blackbodies as calibration sources put at room temperature and liquid nitrogen temperature respectively. Although the chip is configured to be a balanced mixer, we have not measured the performance of the complete system. Because the detailed analysis of the performance of the chip is not the focus of this study, we defer this work until confirming every part of the SoC functions as expected. In this study the four IF outputs from the chip are independently treated without combining with 180-degree IF hybrid. Each output is connected to a cryogenic isolator with 4-8 GHz passband followed by a cryogenic low noise amplifier with a nominal noise temperature of 8 K. One of the four low temperature IF channels is selected with a coaxial switch to be connected to a room-temperature amplifier chain. A power meter and a spectrum analyzer are used to do the power measurement and analysis.

Examples of the IV curves and the noise temperature measurement results are given in Figure 13. Figure 13(a) shows that SIS a1 (a2) and SIS b2 (b1) are bumped at the same level but the mixers at the same polarization are pumped at slightly different levels. The results indicate that the LOs distributed to the two polarization branches by the Y-junction power divider are in good balance, while the following branch-line couplers do not divide the LO equally. The unbalance of the hybrid bridge and its impact on the balanced mixing has not been evaluated yet. Figure 12(b) shows a typical noise temperature measurement result. The minimum noise temperature about 70 K is reasonably good if a nominal loss of 3 dB at the hybrid bridge is taken into account. This indicates that the loss of the transmission line in front of the SIS mixers is reasonably low at 2 mm wavelength as expected.

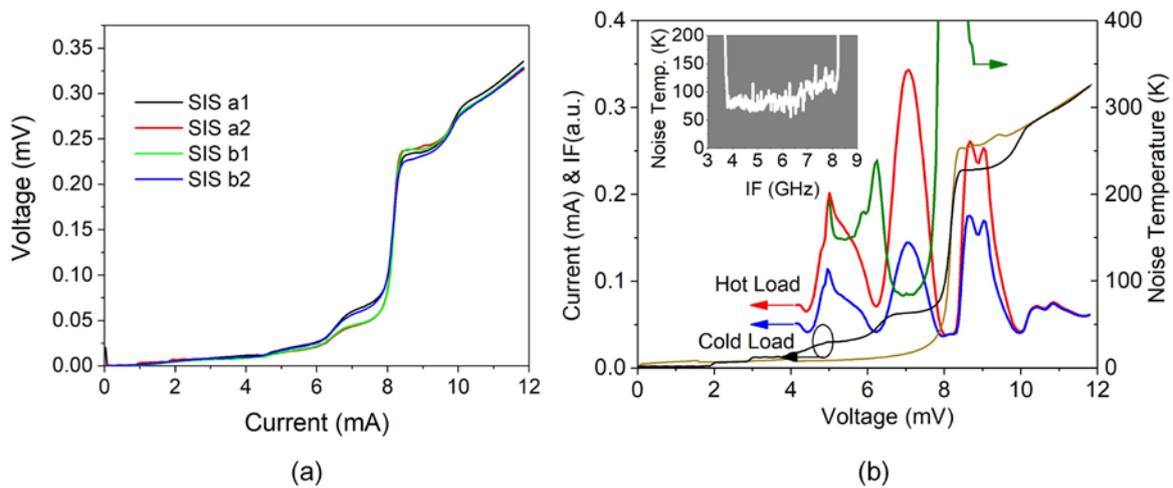

Figure 13. (a) The pumped IV curves of the four SIS junction arrays at LO = 133 GHz. (b) The noise temperature measured as functions of bias voltage and IF at LO = 155 GHz.

The performance of the combination of the planar OMT and the corrugated feed horn is measured with a xyz-θ scanner put outside the vacuum window, on which a CW signal source with a probe feed horn is mounted. The SIS mixer is tuned to receive the CW signal and the output IF is monitored by the spectrum analyzer. The outputs of the four channels were recorded while rotating the scanner (θ) with an example shown in Figure 14(a). A low cross-polarization better than 1% has been confirmed at the measured frequency. By raster scanning the CW source, the co-polarization and cross-polarization patterns were measured, an example of which is shown in Figure 14(b). Except for the irregular shape of the cross-polarization pattern, the cross-polarization level is lower than -23 dB. This low cross-polarization level complies with the requirement for polarization measurement in radio astronomical observation. We found that the irregularity of the cross-polarization pattern is caused by cross talk between the two polarization branches through the Y-junction LO waveguide divider. If the on-chip branch-line coupler is not perfectly matched, part of the signal will appear in the isolated port, where LO enters. The signal goes upstream to the LO probe and enters the LO waveguide. Because the Y-junction waveguide divider has relatively low isolation (-6 dB) between the two outputs, the signal can travel across the Y-junction and enter the other polarization branch. An obvious solution is to change the Y-junction divider to a waveguide branch-line coupler to enhance the isolation between the two polarization circuits.

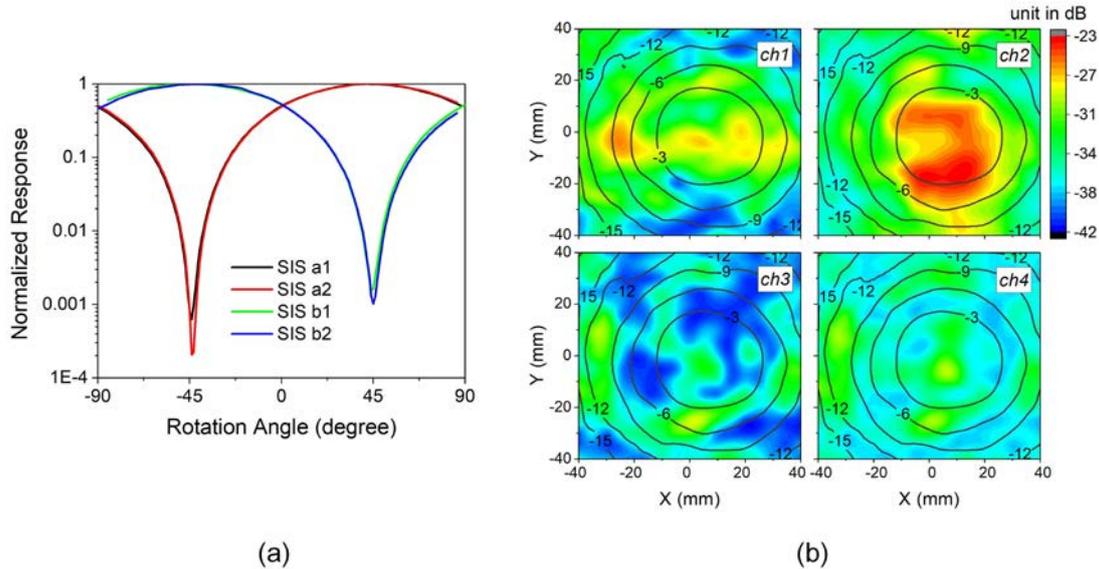

Figure 14. The performance of the planar OMT/corrugated horn combination measured with LO frequency 145 GHz and signal frequency 139 GHz. (a) The responses of the mixers with rotating the scanner; (b) the beam pattern measured from the four IF outputs.

## 5. CONCLUSION

We have introduced a design concept that enables planar integration of waveguide SIS mixer array receiver frontends. This concept can be implemented to build a focal array receiver with very compact profile at mm and possibly extendable to sub-mm wavelengths. Compared with the conventional SIS mixer arrays based on building blocks of single-beam modules, planar integrated arrays are supposed to have much higher reliability because of its high level of integration. This method has the exceptional merit that one can incorporate dual-polarization, balanced mixing and sideband separation into very compact heterodyne arrayed receiver frontend. The feasibility of this concept has been partially demonstrated with a one-pixel prototype model operating at 2 mm wavelength.

## ACKNOWLEDGEMENT

The authors would like to thank Yutaro Sekimoto of ISAS JAXA and Shibo Shu of IRAM for their assistance in the planar OMT design. The authors are also grateful to Keiko Kaneko of NAOJ for her assistance in mixer mount design, to Akihira Miyachi and Kroug Matthias of NAOJ for their helpful discussion in device fabrication, and to Daisuke Iono for his discussion in scientific values of wide FoV observations. The work is supported in part by the Japan Society for the Promotion of Science (JSPS) KAKENHI under Grant Number JP16K13789 and by the National Natural Science Foundation of China under Grants U1331203.